\documentclass[twocolumn,pra,aps,superscriptaddress,showpacs]{revtex4}
\usepackage{xspace,amsmath,amsfonts,amsthm,amssymb,amsbsy,graphicx,color}

\newcommand\bx{\mathbf{x}}
\newcommand\by{\mathbf{y}}
\newcommand\ny{y} 

\newcommand\bu{\mathbf{u}}
\newcommand\bv{\mathbf{v}}

\newcommand\bp{\mathbf{p}}
\newcommand\bcdot{\boldsymbol{\cdot}}

\begin{document}

\newtheorem{theo}{Theorem}
\newtheorem{lemma}{Lemma}

\title{Quantum Feedback Control: \\ How to use Verification Theorems and Viscosity Solutions to Find Optimal Protocols}

\author{Kurt Jacobs}
\affiliation{Department of Physics, University of Massachusetts at Boston, 100 Morrissey Blvd, Boston, MA 02125, USA}
\author{Alireza Shabani}
\affiliation{Department of Electrical Engineering, University of
Southern California, Los Angeles, CA 90089, USA}

\begin{abstract}
While feedback control has many applications in quantum systems, finding optimal control protocols for this task is generally challenging. So-called ``verification theorems'' and ``viscosity solutions'' provide two useful tools for this purpose: together they give a simple method to check whether any given protocol is optimal, and provide a numerical method for finding optimal protocols.  While treatments of verification theorems usually use sophisticated mathematical language, this is not necessary. In this article we give a simple introduction to feedback control in quantum systems, and then describe verification theorems and viscosity solutions in simple language. We also illustrate their use with a concrete example of current interest.
\end{abstract}

\maketitle

\section{Introduction}

In quantum feedback control (QFC), an observer continuously monitors a quantum system~\cite{JacobsSteck06, Brun02}, and uses the information from this measurement, as it is obtained, to control the system by continually modifying the Hamiltonian and/or the observable being measured. Over the last decade the topic of QFC has generated an increasing amount of theoretical~\cite{Belavkin87, Belavkin83, Wiseman95,  Wiseman95b, Yanagisawa98, DJ, Belavkin, DHJMT, DJJ, Thomsen02, Yanagisawa03a, Yanagisawa03b, Geremia03, Bouten04, James04, Bouten05, vanHandel05, Gough05, Bouten05e, Wiseman05, Mirrahimi05-1, Mirrahimi05-2, DHelon06, Dong06, Combes06, Altafini06, Altafini07, Altafini07b, Ticozzi06, Pavon06, vanHandel07, Wang07, Jacobs07c, Ticozzi08, Gough08, James08, Nurdin08, Belavkin08} and experimental~\cite{Smith02, Armen02, Cook07, Higgins07} work. This is due partly to experimental progress in micro- and mesoscopic quantum systems~\cite{Smith02, Armen02, Naik06, Cook07, Higgins07, Niskanen07, Houck07, Schuster07, Plantenberg07}, partly because QFC possesses a wide range of potential applications~\cite{Wiseman95, Wiseman95b, Ahn02, rapidP, Hopkins03, Geremia04, Ralph04,  Sarovar04, Steck04, Steixner05, Yamamoto05, Yanagisawa06, Wiseman06x, Bushev06, Matsukevich06, Mancini06, Katz07, Tian07, Jacobs07, Hill07, Carvalho07, Wilson07, Yamamoto07, Bradley08, Combes08, Wiseman08, Sagawa08}, and partly because of its connection with fundamental questions in quantum mechanics~\cite{DJJ, FJ, Jacobs07c, Sagawa08}.

Unlike in classical feedback control, in QFC the measurement that is part of the feedback loop affects the dynamics of the system~\cite{JacobsSteck06,Brun02}. Even so, QFC is actually contained within the general framework of classical control theory. This is because, even if the measurement in the feedback process is continually modified as the system evolves, quantum systems are merely specific examples of noisy non-linear dynamical systems. As a result, the techniques of classical control theory are usually applicable to quantum systems. However, most of the powerful results of control theory apply only to linear systems, and are therefore not relevant to the majority of quantum feedback control problems. An important exception are the ``verification theorems''~\cite{Fleming75, Boltianskii66, Vinter80, Zhou97, Gozzi05}. These are applicable to feedback control in any dynamical system, and have two uses. The first is to test whether or not  a given feedback protocol is the {\em optimal} protocol for a given task. If one has devised a protocol, by intuitive means or otherwise, it is very useful to be able to check if it is optimal, especially if the protocol admits an analytic solution. Knowing that a protocol is optimal is useful, but so is knowing that it is not --- this tells us that there may yet be hidden and unexpected things to learn about the given problem.  The second use of verification theorems is that they provide a systematic numerical procedure for finding optimal feedback protocols.

The literature on verification theorems is not easily accessible to most quantum physicists, however, because it is written in the jargon of axiomatic probability theory (filtrations, adapted processes and the like). Further, the most widely applicable verification theorem, proved in the last few years, requires the use of ``superderivatives'' and ``viscosity solutions''~\cite{Crandall92}, a subject unfamiliar to most theoretical physicists. Our purpose here is to explain, in a straightforward manner and without technical jargon, how to check if a feedback protocol is optimal, and how to find optimal protocols numerically.  We also give a concrete example of the former, involving an optimal feedback protocol that we reported recently in Ref.~\cite{Shabani08}, and for which the viscosity verification theorem is essential. We note that we were first introduced to verification theorems by the article of Wiseman and Bouten, in which they used these theorems to prove the optimality of a number of quantum rapid-purification protocols~\cite{rapidP,Wiseman08}.  Since the usefulness of verification theorems in QFC is clear, we felt that an accessible article on this topic would useful.

We begin in the following section by introducing the subject of
quantum feedback control itself. While the concepts are simple, it
does require stochastic calculus, and this is unfamiliar to many
physicists. We thus start by discussing this aspect of QFC a little.
We then present the equations of motion for a continuously monitored
system, and show how the effect of feedback can be easily included.
We next show how to quantify the problem of optimizing the feedback
protocol for a given task. Having done this we now know what the
optimization problem is for any given control task, and we can show
how to use the ``classic'' verification theorem to check whether a
feedback protocol is optimal. This is the subject of Sections~\ref{CVT}
and~\ref{vteg}. The classic verification theorem is sufficient so
long as the evolution induced by the feedback protocol has
continuous first and second derivatives. For many protocols the
evolution of the system does not satisfy this condition, however. An
example is a protocol which switches abruptly at some time from one
Hamiltonian to another. In this case the evolution has continuous
derivatives everywhere except at the point(s) where the protocol
switches. To check optimality in this more general case we require a
more general verification theorem. This theorem, while very similar
to the first, requires the use of viscosity solutions. In
Section~\ref{VS} we explain how to calculate a superderivative (and
subderivative), and how to show if a function is a ``viscosity
solution'' of a given differential equation. We can then show how to
use the more general verification theorem to determine the
optimality of a feedback protocol that has  discontinuities, and we
do this in Section~\ref{EVT}. With this we have achieved our main
goal. In Section~\ref{findop} we show how the verification theorems
also provide a numerical method to find optimal feedback protocols,
and in Section~\ref{OCO} we show how, with almost no modification,
the same techniques work for time-optimal control problems.
Section~\ref{conc} concludes with a brief summary.

\section{Quantum Feedback Control}
\label{QFC}

\subsection{Including Feedback in Quantum Mechanics}

The dynamics of an isolated quantum system is given by Schr\"{o}dinger's equation. If we describe the system using a density matrix, $\rho$, then this equation is
\begin{equation}
  \dot{\rho} = -(i/\hbar)[H,\rho] .
\end{equation}
If there is an environment that introduces noise into the system,
then one can often include the effects of this environment by
adding the terms of a Lindblad master equation to the above
Hamiltonian evolution~\footnote{We can do this so long as the environment is memoryless (Markovian).}. The equation of motion for $\rho$ becomes
\begin{equation}
  \dot{\rho} = -(i/\hbar)[H,\rho]  + \gamma \mathcal{D}[c] \rho ,
\end{equation}
where we have defined
\begin{equation}
   \mathcal{D}[c] \rho = 2c^\dagger \rho c - c^\dagger c \rho - \rho c^\dagger c .
\end{equation}
Here $\gamma$ is a rate constant, and $c$ is an operator that depends on the coupling of the system to the environment. To add to the equation of motion the effect of monitoring the system, we simply add another set of terms. Before we do so, however, let us explain exactly what we mean by monitoring.

Monitoring a system means that we obtain a continuous readout of some property of the system. Let us say that we are monitoring the position of an object. This means that in each tiny time-step $dt$, we obtain a little bit of information about this position, $x$. Why a tiny bit? Why not the precise position? To understand this, first recall that all real measurements (classical as well as quantum) have some inaccuracy, which means that the measurement result is $x$ plus some random number (the error). This random number limits the amount of information we have about $x$ (that is, it limits the accuracy to which we can pinpoint $x$ after obtaining the result). Now, no physical measurement process can extract a finite amount if information in an infinitely short time, because the speed of any interaction is necessarily finite.  Thus in a time interval $dt$, the amount of information must also be infinitesimal, which means that the error must tend to infinity as $dt \rightarrow 0$. For a classical measurement the measurement result in a time interval $dt$ is therefore given by
\begin{equation}
   R = x + \varepsilon ,
\end{equation}
where $x$ is the true value of $x$, and $\varepsilon$ must have a bigger variance the smaller
$dt$. It turns out that the quantum equivalent of this is~\cite{JacobsSteck06}
\begin{equation}
   R = \langle X \rangle + \varepsilon ,
\end{equation}
where $X$ is the operator for the observable $x$~\footnote{While
it makes sense that for a classical measurement
of $x$ the result should be the true value plus an error, it is not
so clear why the expression $\langle X \rangle + \varepsilon$ also
makes sense. Nevertheless, it turns out that the two forms are
completely equivalent. It follows that the only way that a
measurement of $x$ can determine the expectation value of $x$
precisely is to determine the true value of $x$ precisely, at which
point the expectation value and the true value are the same.}. It
turns out that if the error is Gaussian, (being by far the most
common measurement error because of the central limit theorem, and
the one we will consider here), the variance of $\varepsilon$ must
be proportional to $1/dt$ (and thus the error is proportional to
$1/\sqrt{dt}$). For further details on the reason for this, we refer
you to~\cite{WienerIntroPaper, JacobsSteck06}. Because of this the
value of $R$ becomes, strictly speaking, infinite as $dt \rightarrow
\infty$. So instead of writing our equations in terms of $R$, we
write them instead in terms of $R dt$, which we will denote by $dr$.
This is a perfectly finite quantity. Se we have
\begin{equation}
   dr = \langle X \rangle dt + g dW ,
\end{equation}
where $g$ is an arbitrary constant determining the amount of the noise, and $dW$ is a Gaussian random variable with mean zero and variance equal to $dt$. Thus in each interval $dt$, $dW$ is picked from the probability density
\begin{equation}
   P(dW) = \frac{1}{\sqrt{2\pi dt}}e^{-dW^2/(2dt)} .
\end{equation}

Now that we know precisely what we mean by a continuous measurement of a physical quantity $x$; we mean a measurement that provides a continuous stream of measurement results $dr(t)$. Now we need to know how the state $\rho$ changes in each time interval as a result of the measurement. In the interval $dt$, the state $\rho(t)$ goes to $\rho(t) + d\rho$, where~\cite{JacobsSteck06}
\begin{equation}
   d\rho = - k \mathcal{D}[X]\rho +  \sqrt{2k}(X\rho + \rho X - 2\langle X\rangle \rho ) dW ,
\end{equation}
where $ k  = 1/(8g^2)$ and as usual $\langle X\rangle = \mbox{Tr}[X\rho]$. Loosely speaking, $k$ gives the rate at which the measurement extracts information about $x$. Here the random increment $dW$ is precisely the {\em same}  increment that appears in the measurement record $dr$. This makes sense: the change in the state of the system in the time interval $dt$ depends upon the measurement result in that time interval. As a result we can also write the above equation for $d\rho$ directly in terms of the measurement record:
\begin{equation}
   d\rho = - k \mathcal{D}[X]\rho +  4k(X\rho + \rho X - 2\langle X\rangle \rho )(dr - \langle X \rangle dt) .
\end{equation}
Normally one would divide $d\rho$ by $dt$ to get the time derivative $\dot{\rho}$, and write the equation of motion for $\rho$ using this derivative. However, $dW/dt$ is rather pathological (just as $r$ is above), so when writing differential equations involving $dW$ we keep them in terms of differentials. Thus the full equation of motion for a noisy system that is also continuously monitored is
\begin{eqnarray}
   d\rho & = & -(i/\hbar)[H,\rho] dt + \mathcal{D}[c]\rho dt  - k \mathcal{D}[X]\rho dt  \nonumber \\
             &  & +  \sqrt{2k}(X\rho + \rho X - 2\langle X\rangle \rho ) dW .  \label{smerec}
\end{eqnarray}
This equation is referred to as a {\em stochastic} equation, because it contains a term that fluctuates randomly.

To be able to manipulate stochastic equations, one needs to use the
rules of {\em stochastic calculus}. This boils down entirely to the rule
$dW^2 = dt$. On first sight this is a very strange relation, because
$dW$ is random, and $dt$ is deterministic. However, note that
$\langle dW^2 \rangle = dt$. Because of this, if one divides any
tiny time interval into many smaller sub-intervals, and adds up the
many increments of $dW^2$, the result is a deterministic quantity in
the limit in which there are infinitely many
sub-intervals~\cite{KJPhD}. The result is that in the continuum
limit it is always true that $dW^2 = dt$. For further details
regarding the manipulation and solution of stochastic equations we
refer you to~\cite{WienerIntroPaper,KJPhD,JacobsSteck06,Brun02}.

Now that we have the evolution equation for a monitored system, we want to include feedback in the dynamics. This is very simple. All we do is specify that the observer can change the Hamiltonian of the system, $H$, and also if we wish, the measured observable $X$, and allow $H$ and $X$ to be some function of the measurement results. Specifically, $H$ and $X$ at time $t$ are allowed to be any function of the measurement results, $dr(t)$, obtained up until that time. We can write $H$, for example, as $H(\mathcal{F}(t))$, with $\mathcal{F}(t) = \int_0^t f(t,t') dr(t')$ where $f(t,t')$ is an arbitrary function of its arguments.  Because the density matrix at time $t$ completely determines the future behavior of the system (given the system Hamiltonian), all optimal feedback protocols can be obtained by making $H(t)$ and $X(t)$ functions of $\rho(t)$ (and possibly the current time, $t$). The observer obtains $\rho(t)$ simply by using the measurement results to solve Eq.(\ref{smerec}) from the initial time up until time $t$. Thus we choose $H = H(t,\rho)$ and $X = X(t,\rho)$. The equation for the evolution of $\rho$, including feedback, is then simply
\begin{eqnarray}
   d\rho & = & -(i/\hbar)[H(t,\rho),\rho] dt - k \mathcal{D}[X(t,\rho)]\rho dt   \notag \\
             &  &  +  \sqrt{2k}[X(t,\rho) \rho + \rho X(t,\rho) - 2\langle X(t,\rho) \rangle \rho ] dW \notag \\
             &  & + \mathcal{D}[c]\rho dt    .  \label{fullsme}
\end{eqnarray}
The problem of feedback control is to determine $H(t,\rho)$ and/or $X(t,\rho)$ in order to achieve a desired evolution as closely as possible in the presence of noise.

\subsection{The Goal of Control}
In feedback control the goal is usually one of three things:  to arrive at a desired state at a given time $T$; to produce an evolution that is as close as possible to some desired evolution; or to reach a given state as quickly as possible. The first two are called ``finite-horizon" control, and the third ``time-optimal" control. We will consider the first two of these when introducing the verification procedures. These procedures can be applied equally well to the third, with minor modification, and we discuss this in Section~\ref{OCO}.

The reason that the first two control objectives fall in the same
category is that the first is merely a special case of the second:
in the former we require the system to be close to a specific pure
state at a given ``horizon time" $T$; in the latter we require
that the system be as close as possible to some given (time
dependent) pure state, $|\psi(t)\rangle$, at {\em every} time. This
desired state is called the {\em target state}, or the target
evolution. (Note that there is little point in choosing the target
state to be mixed - mixing merely represents a lack of knowledge
regarding the state, and this is not beneficial in a control
setting.)

Now consider the second control objective. To define precisely how well a control protocol achieves the objective, we need to choose a measure of the distance from one  quantum state to another to quantify what we mean by the system being {\em close} to the target state. There are a number of measures we can use, such as the fidelity~\cite{FuchsPhD}, the distinguishability~\cite{Fuchs98}, or the quantum relative entropy~\cite{Vedral97}, to name but three. If we choose the fidelity then we want our control protocol to minimize
\begin{equation}
   F(t) = 1 - \langle\psi(t)|\rho(t)|\psi(t)\rangle.
\end{equation}
Since this is a function of time, to obtain a single quantity to
minimize we must combine $F(t)$ at all the different times into a
single number. To use verification theorems, and indeed the majority
of results of optimal control theory, we must combine the $F$'s in a
simple fashion: the function to minimize must be merely a weighted
sum of the $F$'s at different times. Since time is continuous, this
means that the function to minimize is $J = \int_0^T \alpha(t)F(t)
dt$, for some $\alpha(t)$ and final time $T$. This means that the
function to minimize is simply an integral over time of a function
of the state $\rho$ at each time. Fortunately this form is quite
reasonable and well-motivated, and is rather general. For example,
we can weight different times differently in the integral if some
are more important that others. If the final time is especially
important, we can give it extra weight by including it by itself in
addition to the integral (that is, place a delta function weighting
at time $T$). We can also include more general functions of the
state, since we can place any function of the form $L(\rho(t),t)$ in
the integral. For example, if we just want to control the
expectation value of a particular observable then $L$ can be a function of
this expectation value. Or we could choose $L$ to be the von Neumann
entropy of the state, as another example.

There is one more crucial thing to add. If we could both extract information from the system infinitely fast, and apply infinite forces to the system (control Hamiltonians that induce infinitely fast dynamics) then we could easily make the system follow any target evolution exactly, even in the presence of environmental disturbance and other noise. Feedback control is only non-trivial if the measurement rate and/or feedback forces are constrained, which they always are in real applications. There are two ways to place constraints on the control inputs and the measurement rate. One is simply to place a fixed upper bound on them, and attempt to minimize $J$ under these constraints. The alternative is to include a function of the measurement rate and feedback Hamiltonian in $L$. This means that in minimizing $J$ we are trying to find the protocol that gives us the best evolution, while at the same time keeping the control forces as small as possible. The size of the feedback forces, and the rate at which we extract information, are referred to as the ``costs" of the control protocol. Because one often includes these in the functional to minimize, $J$, this functional is usually called the {\em cost}.

Finally, since the system is driven by noise, the actual cost (effort) expended by the controller on a given run will depend on the specific values that the noise takes for that run. These specific values are called the {\em noise realization}. Since the actual cost will vary from run-to-run, it is sensible to have the control protocol minimize the {\em average} value of $J$, where the average is over all noise realizations.

To summarize succinctly the above discussion, a feedback control problem is defined by the equation of motion of the system we are trying to control, and a cost of the form
\begin{equation}
   J = \left\langle \int_0^T \!\! L\left[\rho(t),t,H_{\mbox{\scriptsize fb}}(t),X(t)\right] dt + M(\rho(T)) \right\rangle_{\mbox{\scriptsize n}},
   \label{cost}
\end{equation}
that we choose based on our control objective, and that we want to
minimize. Here $H_{\mbox{\scriptsize fb}}$ is the part of the Hamiltonian
that gives the forces applied by the controller (the ``control
Hamiltonian"), and we have separated
out the contribution to the cost of the final state, calling this
$M(\rho(T))$. It is useful to do this because in many control
problems it is, in fact, only the final state that is important, so
that one sets $L=0$. The angle brackets with the subscript ``n", $\langle \cdots \rangle_{\mbox{\scriptsize n}}$, indicate that the value of the integral is averaged over all possible noise realizations.

\section{The Classic Verification Theorem}
\label{CVT}

We now explain how to use the ``classic'' verification theorem of optimal control theory to determine whether a given protocol is the optimal one.

\subsection{First, a few definitions...}
We will now move to a new notation. This is to make the presentation easier, and to employ the same notation as used by most control theory texts. We write the general dynamical equation for the system we are trying to control as
\begin{equation}
   d\mathbf{x} = \mathbf{A}(t,\mathbf{x},\mathbf{u}(\mathbf{x},t)) dt + B(t,\mathbf{x},\mathbf{u}(\mathbf{x},t)) \mathbf{dW} .
\label{sys-dy}
\end{equation}
Here the state of the system is given by the vector $\mathbf{x}$, and the control inputs to the system are denoted by the vector $\mathbf{u}$ (which is usually a function of the current state, $x$)~\footnote{So long as the control at time $t$ does not depend explicitly on the state of the system at {\em earlier} times, the control is referred to as {\em Markovian}. In fact, so long as the current state is the observers complete state-of-knowledge determined from the measurement record, in theory there is never any need to have the current control depend upon the system at earlier times, since the complete future evolution is entirely determined by the current state.}. Thus for quantum feedback control, the vector $\mathbf{x}$ is the vector of the elements of $\rho$:
\begin{equation}
   \mathbf{x}   \Longleftrightarrow \rho . 
\end{equation}

The vector $u$ is the set of all parameters that we can vary. If we write the control Hamiltonian as
\begin{equation}
   H_{\mbox{\scriptsize fb}}(t) =  \mu_1(t) H_1 + \mu_2(t) H_2 + \cdots + \mu_i(t) H_i + \cdots , 
\end{equation}
where the $\mu_i$ are real numbers, then $\mathbf{u}$ is the set of these $\mu_i$, along with a set of numbers that determine the observable we choose to measure, $X(t)$. If we write $X$ as
\begin{equation}
   X = U D U^\dagger ,
\end{equation}
where $U$ is unitary and $D$ is diagonal, then $X$ is determined in the most general case by the diagonal elements of $D$, and the $N(N+1)/2$ angles $\theta_k$ that parametrize an $N$-dimensional unitary~\cite{Dita03}.  Thus
\begin{equation}
   \mathbf{u}   \Longleftrightarrow  \{ \mu_i , D , \theta_k \} .
\end{equation}

The vector $\mathbf{A}$ is any vector-valued function of $\mathbf{x}$, $\mathbf{u}$ and $t$, and gives the deterministic dynamics of the system. It is determined by the Hamiltonian $H$ and the noise operator $c$. (Yes -- the noise on the system actually gives deterministic motion, because we are considering the evolution of the density matrix. It is only the measurement that makes the evolution stochastic.) The matrix-valued function $B$ gives the stochastic part of the evolution. It is determined by the measured observable $X(t)$. The vector $\mathbf{dW}$  is a vector of mutually independent Wiener noises. As usual we will take the initial time to be $t = 0$, denote the initial state of the system as $x(0) = \mathbf{x}_0$, and we will call the final (horizon) time $t=T$.

The function $\mathbf{u}(x(t),t)$ completely defines the control protocol. It tells us what control inputs (what control Hamiltonian) to apply at any time $t$, for every possible state of the system at that time. Thus $\mathbf{u}(x(t),t)$ {\em is} the control protocol.

The control inputs, $\mathbf{u}$, are usually subject to limits.  In general these limits can be defined by any closed region in the vector space in which $\mathbf{u}$ lives. The simplest set of conditions would be
 \begin{equation}
  \mathbf{u}_{\mbox{\scriptsize low}}  \leq  \mathbf{u}(\mathbf{x},t)  \leq  \mathbf{u}_{\mbox{\scriptsize high}} .
\end{equation}
Note that the region (and thus $ \mathbf{u}_{\mbox{\scriptsize low}} $ and $ \mathbf{u}_{\mbox{\scriptsize high}} $) is not a function of $t$ or $\mathbf{x}$; there is only one region that bounds the values control inputs for the duration of the control. The procedure we will describe is general enough, however, to handle a limiting region that varies with time.

As described above, the control objective is to minimize a {\em cost}, $J$.  Using our new notation, this cost is
\begin{equation}
  J =  \left\langle \int_0^T \!\! L(\mathbf{x},\mathbf{u}(\mathbf{x},s),s) ds + M(\mathbf{x}(T))  \right\rangle_{\mbox{\scriptsize n}} ,
\end{equation}
where $L$ and $M$ are functions that we choose. For the verification
procedure, we have to introduce one more quantity, called the {\em
cost function}, $C(\mathbf{x},t)$. This is defined as
the value of $J$ over the interval $[t,T]$, given that the system is
at state $\mathbf{x}$ at time $t$:
\begin{equation}
  C(\mathbf{x},t) =   \left\langle \int_t^T \!\! L \; ds + M(\mathbf{x}(T))  \right\rangle_{\mbox{\scriptsize n}} . 
\end{equation}
Note that $C(\mathbf{x},t)$ is the average cost that
will have to be paid over the time remaining, given we have reached
time $t$. For this reason it is often called the ``cost-to-go".

\subsection{The verification procedure}
To determine whether a given control protocol, $\mathbf{u} = \mathbf{f}(\mathbf{x},t)$ is optimal, one performs the following four steps:

\textbf{1}. Integrate the equations of motion of the system to
calculate the cost function, $C(\mathbf{x}(t),t)$, for
this protocol.

\textbf{2}. Check that $C$ satisfies two continuity conditions. These are that
\begin{equation}
        \frac{\partial C}{\partial t}  \;\;\;\;\; \mbox{and} \;\;\;\;\;  \frac{\partial^2 C}{\partial \mathbf{x}^2}
\end{equation}
are continuous. Here $\partial^2 C/\partial \mathbf{x}^2$ denotes the matrix of second derivatives of $C$. 

\textbf{3}. Determine whether or not $C(\mathbf{x},s)$ satisfies the following differential equation (called the Hamilton-Jacobi-Bellman (HJB) equation):
\begin{equation}
  \frac{\partial C}{\partial t} = \max_{\mathbf{v}} \left[ G\left( \mathbf{x}, \mathbf{v}, \frac{\partial C}{\partial \mathbf{x}}, \frac{\partial^2 C}{\partial \mathbf{x}^2} \right) \right] , 
  \label{HJB}
\end{equation}
with the final condition $C(x,T) = M(x(T))$~\footnote{The terminal condition on the HJB equation, $C(x,T) = M(x(T))$, is always satisfied by the cost function, since this condition follows immediately from its definition.}.

We will give the function $G$ below. Before we do so, we note that the maximum is taken over the allowed values of a vector $\mathbf{v}$. The allowed values of $\mathbf{v}$ are those of $\mathbf{u}$: $\mathbf{v}$ can only take values in the region that bounds the values of $\mathbf{u}(\mathbf{x},t)$. Note that because $G$ is a function of $t$ and $\mathbf{x}$, one must maximize $G$ separately at each time and at each value of $\mathbf{x}$. Because of this, the value of $\mathbf{v}$ that maximizes $G$ will in general be different at different values of $t$ and $\mathbf{x}$. Thus the $\mathbf{v}$ that maximizes $G$ is in general a function of $t$ and $\mathbf{x}$: $\mathbf{v}_{\mbox{\scriptsize max}}(\mathbf{x},t)$.

Now we come to the function $G$. This is
\begin{eqnarray}
  G & = & - \frac{1}{2} \mbox{Tr} \left[ B^{\dagger}(t,\mathbf{x},\mathbf{v}) \frac{\partial^2 C}{\partial \mathbf{x}^2} B(t,\mathbf{x},\mathbf{v})  \right]  \nonumber \\
    & & -  \mathbf{A}  \bcdot \frac{\partial C}{\partial \mathbf{x}} - L(t,\mathbf{x},\mathbf{v}) .
\end{eqnarray}
Note that $\mathbf{A}$ and $B$ are respectively the vector and matrix that appear in the equation of motion for the system, $\partial C/\partial \mathbf{x}$ is the vector of first derivatives of $C$, and $\partial^2 C/\partial \mathbf{x}^2$ is the matrix of second derivatives. To check that $C$ satisfies the differential equation given by Eq.(\ref{HJB}), the only potentially tricky thing is determining what the maximum of $G$ is for each $t$ and $x$ --- however, this can be fairly easy, depending on the system.

\textbf{4}. Check that the $\mathbf{u}(\mathbf{x},t)$ for your protocol maximizes $G$ (if this maximum is unique, then this also means that $\mathbf{u}(\mathbf{x},t) = \mathbf{v}_{\mbox{\scriptsize max}}(\mathbf{x},t)$).



\section{Using the Verification Theorem: An Example}
\label{vteg}

We now apply the verification theorem to a concrete example. This example was solved in Ref.~\cite{Shabani08}, but without the details of the analysis. The problem involves feedback control of a 3-state quantum system (otherwise known as a qutrit). The purpose is to prepare the system in a given pure target state at time $T$, but it is only the measurement strength $k$ that is bounded; it is assumed the feedback Hamiltonian available to the controller generates evolution that is much faster than the measurement strength, and the noise in the system. It is also assumed that the controller has the ability to implement any Hamiltonian for the qutrit.

Because of the strong feedback Hamiltonian, the problem becomes one of maximizing the largest eigenvalue of the density matrix. The reason for this is that, at any desired time, the controller can quickly evolve the system (that is, apply a unitary operator) so that the eigenvector corresponding to the largest eigenvalue is equal to the target state. For a given density matrix, this unitary operation maximizes the probability that the system is in the target state, and is thus the optimal application of the Hamiltonian part of the feedback. In this case the probability that the system is in the target state is equal to the largest eigenvalue of the density matrix. The feedback control problem in this case is therefore to choose the measured observable, $X(t)$, as the system evolves, so as to maximize the largest eigenvalue of the density matrix at time $T$. 

Since the feedback Hamiltonian is very large, it can also be used to continually rotate the eigenvectors of the density matrix so the density matrix eigenbasis remains constant with time. The result of this is that, when we express $X$ in terms of this eigenbasis, the equations of motion for the eigenvalues of the density matrix are decoupled from those of the eigenvectors, and so we only need to consider the motion of the eigenvalues to find the optimal control protocol. 

In addition to the restriction on the measurement strength, we restrict the observable $X$ to having equi-spaced eigenvalues. Since the measurement part of the master equation is invariant under the transformation $X \rightarrow \alpha I$, where $\alpha$ is a real number, $X$ can be taken to be traceless without loss of generality. As a result, the measured observable is of the form
\begin{equation}
  X = U        \left(
            \begin{array}{ccc}
                a & 0  & 0  \\
                   0    & 0  & 0  \\
                    0   &  0 &   -a
            \end{array}
                   \right)                      U^\dagger  ,  \label{Xcon}
\end{equation}
where $U$ is any unitary. 

Before solving the problem, we specialize to the ``regime of good control".
This is the regime in which the probability that the system is in the
target state, $P$, is close to unity~\cite{Jacobs07c, Shabani08}. This means
that $P = 1 - \Delta$, where $\Delta \ll 1$. One can then simplify the dynamics
of the density matrix by expanding to first order in $\Delta$. Recall that 
the goal of the feedback in this case is to optimize the largest value of the 
density matrix, and this problem is completely defined by the dynamics of the 
eigenvalues alone. 

We now denote the largest eigenvalue by $\lambda_0$, the second largest by $\lambda_1$, and the smallest by $\lambda_2$. Since $\sum_{i}\lambda_i = 1$, we have
only two independent dynamical variables, $\lambda_1$ and $\lambda_2$.
In the regime of good control, under a measurement of $X$, the equations
of motion are~\cite{Shabani08} 
\begin{eqnarray}
    d\lambda_1 & = & - 8 \left[ |X_{10}|^2 \lambda_1 - |X_{21}|^2 \frac{\lambda_1 \lambda_2}{\lambda_1 - \lambda_2} \right] dt \nonumber \\
                        &   & + \sqrt{8} (X_{00} - X_{11}) \lambda_1 dW  \label{eqem1} \\
    d\lambda_2 & = & - 8 \left[ |X_{20}|^2 \lambda_2 + |X_{12}|^2 \frac{\lambda_1 \lambda_2}{\lambda_1 - \lambda_2} \right] dt \nonumber \\
                        &   & + \sqrt{8} (X_{00} - X_{22}) \lambda_2 dW \label{eqem2} ,
\end{eqnarray}
where the matrix elements of $X$ are those in the eigenbasis of the density matrix
at the current time. Note that the task is to choose $X(t)$ so as to maximize $\lambda_0(T)$, which means minimizing $\Delta(T) = \lambda_1(T) + \lambda_2(T)$.

Our candidate for the optimal control protocol, suggested in~\cite{Shabani08}, is
to choose $X$ at each time so that, in the eigenbasis of the density matrix, it is
\begin{equation}
  X = \left(
            \begin{array}{ccc}
                     0 & a  & 0  \\
                     a & 0  & 0  \\
                     0 &  0 &  0
            \end{array}
        \right) .                             \label{Xmax}
\end{equation}
This is suggested as the optimal protocol {\em only} up until the point at which $\lambda_1$ has
been reduced to the value of $\lambda_2$. We now use the procedure described in the previous section to determine whether or not this protocol is optimal. (If you really want to burn this procedure into your brain, you can stop reading now and try it.)

To begin we need to obtain the cost function. For this we need the time evolution of the thing we want to minimize (the cost), $\Delta$, under the suggested protocol.  Under the protocol, the only non-zero elements of $X$ are $X_{01} = X_{10} = a$, and the equations of motion become
\begin{eqnarray}
    d\lambda_1 & = & - 8 a^2 \lambda_1 dt , \\
    d\lambda_2 & = & 0 .
\end{eqnarray}
These are easy to solve, with the result that $\Delta(t) = \lambda_1(0) e^{-8 a^2 t} + \lambda_2(0)$. Recall that the cost function, $C(\lambda_1,\lambda_2,t)$, is the value of $\Delta$ at the final time, $T$, given that we {\em started} with the values $\lambda_1$ and $\lambda_2$ at time $t$.  So this is
\begin{equation}
   C(\lambda_1,\lambda_2,t) =  \lambda_1 e^{-8 a^2 (T-t)} + \lambda_2 .
   \label{W1}
\end{equation}
Note that all the derivatives of $C$ are continuous, so the continuity conditions for the verification theorem are satisfied.

Now we have to calculate the function $G$. The ingredients for this are the equations of motion, Eqs.~(\ref{eqem1}) and (\ref{eqem2}), and the first and second derivatives of $C$ with respect to $\lambda_1$ and $\lambda_2$. Calculating these derivatives we find that
\begin{eqnarray}
   G(X) & = & 8\left[ \lambda_1 |X_{10}|^2 e^{-8a^2 (T-t)} + \lambda_2 |X_{20}|^2\right] \nonumber \\
           &  & + 8 \left( \frac{\lambda_1 \lambda_2}{\lambda_1 - \lambda_2} \right) |X_{21}|^2  (1 -  e^{-8a^2 (T-t)} ) . 
\end{eqnarray}
We must now find the maximum of $G(X)$ over all $X$, under the constraint Eq.(\ref{Xcon}).  Thus we must maximize $G$ over all unitaries $U$. It is important to note that the $a$ that appears in $G(X)$ above it not part of the optimization; this has already been fixed by the protocol. To perform the maximization, we note first that we are only considering times $t$ such that $\lambda_1 e^{-8a^2 t} \geq \lambda_2$. Because of this, we can show that $G$ will be maximized if we maximize
\begin{eqnarray}
   F(X) = 8 \lambda_2 \left[ \eta |X_{10}|^2 +  |X_{20}|^2 + \xi |X_{21}|^2 \right] ,
\end{eqnarray}
where $\eta$ and $\xi$ are constants satisfying $\eta > 1 > \xi$.
(Specifically $\eta = \lambda_1 e^{-8a^2 (T-t)}/\lambda_2$, and $\xi
= \eta [e^{8a^2 (T-t)} - 1]/[\lambda_1/\lambda_2 - 1]$.) We do this
maximization over $U$ numerically using Matlab's fminsearch
function. (This function uses the Nelder-Mead direct search
algorithm.) This shows that the maximum is obtained when $X$ is
given by Eq.(\ref{Xmax}). Substituting this into $G(X)$ we have
\begin{eqnarray}
  \max_{U} G(X) =  \lambda_1 8 a^2 e^{-8a^2 (T-t)}  .  \label{cvtRHS}
\end{eqnarray}
Calculating the time derivative of $C$ we have
\begin{eqnarray}
 \frac{\partial C}{\partial t} =  \lambda_1 8 a^2 e^{-8a^2 (T-t)} .  \label{cvtLHS}
\end{eqnarray}
Eqs.~(\ref{cvtRHS}) and (\ref{cvtLHS}) are equal. Thus the cost function satisfies the Hamilton-Jacobi-Bellman equation (Eq.(\ref{HJB})), and so the protocol is optimal.

Recall that we have only tested the protocol up until the point at which the lowest two eigenvalues equalize. If $t$ is the starting time, and $T$ the final time, then this means that
\begin{equation}
    T \leq t + \frac{\ln(\lambda_1/\lambda_2)}{8a^2} .
\end{equation}

\section{Viscosity Solutions}
\label{VS}

The purpose of ``viscosity solutions'' (a branch of the theory of differential equations) is to allow one to define a continuous function to be a solution of a second order differential equation, even though the function does not have well-defined derivatives. The reason that this subject is useful to us here, is that optimal feedback protocols often produce cost functions ($C(\lambda_1,\lambda_2,T)$ in the previous section) that are not differentiable at some (usually small, finite) set of points. It turns out that such cost functions are viscosity solutions of the HJB equation, and as a result one can check their optimality by using a verification theorem that is very similar to the classic verification theorem in the previous section. So it turns out yet again that an initially rather odd-seeming branch of mathematics has a direct application.

To use the more general verification theorem we first have to know how to show if something is a viscosity solution to a given DE. To do this we need two new definitions, that of a {\em superderivative} (also called a {\em superjet}) and a {\em subderivative} (or {\em subjet}). To begin with, note that if a function $f(\by)$ is twice-differentiable at point $\by_0$, then
\begin{equation}
  f(\by) -  f(\by_0) =  \bp\cdot (\by-\by_0) + \frac{(\by-\by_0)^\dag Q(\by-\by_0)}{2}
\end{equation}
to second order in $\by-\by_0$, where
\begin{equation}
   \bp = \frac{\partial f}{\partial \by}
\end{equation}
is the vector of first derivatives, and $Q$ is the matrix of second derivatives:
\begin{equation}
   Q_{ij} = \frac{\partial^2 f}{\partial y_i \partial y_j} , \;\;\;\; \mbox{or}  \;\;\;\;\;\; Q = \frac{\partial^2 f}{\partial \by^2} .
\end{equation}
We now define the superderivative, $J^{+}(\by_0)$ of a function at the point $\by_0$ as the set of all vectors $\bp$ and matrices $Q$ such that
\begin{equation}
  f(\by) -  f(\by_0) \leq  \bp\cdot (\by-\by_0) + \frac{(\by-\by_0)^\dag Q(\by-\by_0)}{2}
\end{equation}
to second order in $\by - \by_0$. The subderivative, $J^{-}(\by_0)$ is similarly defined by replacing the $\leq$ with a $\geq$. In feedback control, the functions one deals with are usually twice-differentiable everywhere except at a small number of points. So as an example, let us calculate the super- and subderivatives of the function
\begin{eqnarray}
  f(\ny) = f_-(\ny) & = & \ln \ny  , \;\;\;\;\;\;\;\;\;\;\;  0 < \ny \leq 1 \\
  f(\ny) = f_+(\ny) & = & (\ny-1)^2 ,  \;\;\;\;\;\;\;\;\;\;\;   \ny \geq 1 ,
\end{eqnarray}
at the point $\ny=1$ (where it is not differentiable). To do this we
first expand $f_\pm$ about the point $\ny = 1$, using the fist two
terms of the Taylor series. This gives
\begin{eqnarray}
   f_-(\ny) - f_-(1) & = &   \Delta \ny - (\Delta \ny)^2  , \\
   f_+(\ny) - f_+(1) & = &  2\Delta \ny^2   ,
\end{eqnarray}
to second order in $\Delta \ny = \ny - 1$. So for the real numbers $p$ and $Q$ to be in the superderivative set, $J^{+}(1)$, we need them to satisfy the following two equations simultaneously:
\begin{eqnarray}
 \Delta \ny - \Delta \ny^2  & \leq &   p \Delta \ny + Q (\Delta \ny)^2 ,  \;\;\;\;\;  \Delta \ny < 0 , \label{jplus1} \\
   2 \Delta \ny^2 & \leq &    p \Delta \ny + Q (\Delta \ny)^2 ,  \;\;\;\;\;  \Delta \ny > 0 . \label{jplus2}
\end{eqnarray}
Note that the second order terms are irrelevant unless the first order terms on each side are equal. Examining equation Eq.(\ref{jplus2}) we see that the condition is satisfied so long as $p>0$, or $p=0$ and $Q\geq2$. Examining equation Eq.(\ref{jplus1}), the condition is satisfied when $p\leq1$ (remember that $\Delta \ny$ is {\em negative} for Eq.(\ref{jplus1})), or $p=1$ and $Q\geq -1$. Since (\ref{jplus1}) and $(\ref{jplus2})$ must be satisfied simultaneously, the superderivative is the set
\begin{equation}
    J^+(1) = \left\{
\begin{array}{rcl}
 0 < p < 1, & & Q \in (-\infty,\infty)     \\
  p =0, &   &  Q\geq 2 \\
  p = 1, &  & Q \geq -1 .
\end{array}
\right.
\end{equation}
For the real numbers $p$ and $Q$ to be in the subderivative, $J^{-}(1)$, they must satisfy
\begin{eqnarray}
 \Delta \ny - (\Delta \ny)^2  & \geq &   p \Delta \ny + Q (\Delta \ny)^2 ,  \;\;\;\;\;  \Delta \ny < 0  \label{jneg1} \\
   2 (\Delta \ny)^2 & \geq &    p \Delta \ny + Q (\Delta \ny)^2 ,  \;\;\;\;\;  \Delta \ny > 0 .  \label{jneg2}
\end{eqnarray}
From Eq.(\ref{jneg1}) we find that $p \geq 1$, and from Eq.(\ref{jneg2}) that $p \leq 0$. Since these cannot both be true at once, $J^{-}(1)$ is the empty set:
\begin{equation}
    J^-(1) = \emptyset .
\end{equation}

Now that we know what superderivatives and subderivatives are ,we can define a {\em viscosity solution} to a differential equation. We first write our differential equation in the form
\begin{equation}
   F\left(\by,f,\frac{\partial f}{\partial \by},\frac{\partial^2 f}{\partial \by^2} \right) = 0 .
   \label{pde}
\end{equation}
A continuous function $f(\by)$ is a viscosity solution of this PDE if it is true that
\begin{eqnarray*}
  F(\by,f(\by),\bp,Q) &\geq& 0 , \;\; \mbox{whenever}  \;\;  (\bp,Q) \in J_+(\by) \\
  F(\by,f(\by),\bp,Q) &\leq& 0 , \;\; \mbox{whenever}  \;\;  (\bp,Q) \in J_-(\by) .
\end{eqnarray*}
If the function $f$ is a viscosity solution of a PDE at a given point, and its derivatives exist at this point, then it is a solution in the normal sense. Therefore, to determine if a function is a viscosity solution to a PDE, one first checks that it is a solution to the PDE in all regions where its first two derivatives are defined. Then one calculates the super- and subderivatives of $f$ at those points where its derivatives are not defined, and checks that $f$ is a viscosity solution at those points.

\section{The Enhanced Verification Theorem}
\label{EVT}

The more general verification theorem, due to Zhou and
collaborators~\cite{Zhou97,Gozzi05}, states that $\bu(\bx,t)$ is an
optimal feedback protocol so long as the following conditions are
satisfied by the cost function,
$C(\bx(t),t)$~\footnote{We note that in their proof of
the enhanced verification theorem, the authors of~\cite{Gozzi05}
took the equation of motion for the system to be real, while the
quantum SME is complex. This should not cause any problem, because a
complex vector differential equation can always be written as a real
vector differential equation with twice the number of variables.
However, to be sure we checked with F. Gozzi, who confirmed that the
proof applies to complex variables without change.}:

\textbf{1}. The cost function must be a viscosity solution of the
HJB equation, and it must be true that $C(\bx(T),T) =
M(T)$. To determine whether these conditions are satisfied, one
first writes the HJB equation in the form
\begin{equation}
    -\frac{\partial C}{\partial t} + \max_{\bv} \left[ G \left(\bx,\bv,\frac{\partial C}{\partial \bx},\frac{\partial^2 C}{\partial \bx^2} \right) \right] = 0 .
    \label{HJB2}
\end{equation}
If we set the vector $\by$ in Eq.(\ref{pde}) to $\by = (t,x_1,\ldots,x_N)$, then the above HJB equation is exactly of the form given in Eq.(\ref{pde}). Because the HJB equation does not depend on any second derivatives that include $t$, to determine whether $C$ is a viscosity solution, we do not need to consider these second derivatives when calculating the super- and subderivatives. Thus, for the purposes of solving the HJB equation, the superderivative at $\bx_0$ and time $t_0$, $J^+(\bx_0,t_0)$, is the set of values of $(q,\bp,Q)$ such that
\begin{equation}
  C(\bx,t) -  C(\bx_0,t_0) \leq  q\Delta t + \bp\cdot \mathbf{\Delta x} + \mathbf{\Delta x}^\dag Q\mathbf{\Delta x} , 
\end{equation}
where $\Delta t = t - t_0$ and $\mathbf{\Delta x} = \bx - \bx_0$. Naturally the subderivative is defined in the same way, with the inequality reversed. Armed with these super- and subderivatives one uses the procedure described in the previous section to determine if $C$ is a viscosity solution to the HJB equation.

\textbf{2}. Check that $\bu(t)$ maximizes $G$.

\textbf{3}.  For each $t$ there must be a $q(t)$, $\bp(t)$ and $Q(t)$, such that:
\begin{equation}
   (q(t),\bp(t),Q(t)) \in J^+(\bx(t),t)
\end{equation}
and
\begin{equation}
  q(t) = G \left(\tilde{\bx}(t),\bu(\tilde{\bx}(t),t), \bp(t), Q(t) \right) ,
\end{equation}
where $\tilde{\bx}(t)$ is the evolution of the system under our feedback protocol $\bu(\bx,t)$. The reason that we no longer need the maximization in this equation, is because $C$ is a viscosity solution to the HJB equation (Eq.(\ref{HJB2})), and $\bu(t) = \bu(\tilde{\bx}(t),t)$ maximizes $G$. This means that setting $\bx = \tilde{\bx}(t)$ and $\bv = \bu(\tilde{\bx}(t),t)$ already maximizes $G$.

\section{Using the Enhanced Verification Theorem: An Example}
\label{evteg}
For this example we consider the same control problem as the previous example (Sec.~\ref{vteg}), but we consider the problem for all evolution times, not merely up until the two smallest eigenvalues have equalized. From this point onwards, the protocol suggested in Ref.~\cite{Shabani08} involves rapidly switching the measured observable between $X$, given by Eq.(\ref{Xmax}), and the observable
\begin{equation}
  X_2 = \left(
            \begin{array}{ccc}
                     0 &  0 & a  \\
                     0 &  0 & 0  \\
                     a &  0 & 0
            \end{array}
        \right)   .                        \label{Xmax2}
\end{equation}
This is equivalent to measuring $X$ while rapidly switching the eigenvectors of the eigenvalues $\lambda_1$ and $\lambda_2$. In the limit of fast switching, the equations of motion of both eigenvalues are identical, and given by
\begin{equation}
  \dot{\lambda}_i = -4 a^2 \lambda_i , \;\;\;\;\;\; i = 1,2 .
\end{equation}
Starting at time $t$, and evolving the eigenvalues using the original protocol up until time $t + \tau$, where $\tau = \ln(\lambda_1/\lambda_2)/(8a^2)$, and then using the second protocol from that time until time $T$, we find that, for $T > t + \tau$, the cost function is
\begin{equation}
   C^-(\lambda_1,\lambda_2,t) =  2 \sqrt{\lambda_1\lambda_2} e^{-4 a^2 (T-t)}.
       \label{W2}
\end{equation}
For times $T < t + \tau$, the cost function is still given by Eq.(\ref{W1}), and we will call this $C^+$.

The cost function is continuous everywhere, but is no longer differentiable at $t = T - \tau$. Thus we cannot use the classic verification theorem on the protocol, but must use the enhanced verification theorem. This requires that we show that $C$ is a solution to the HJB equation on the intervals in which it has continuous derivatives, and that it is a viscosity solution at the point $t = T - \tau$.  We already know that it satisfies the HJB equation on the interval $t \in [T-\tau,T]$ (that is, when the evolution time is less than $\tau$). So we examine next the interval $t \in [0, T-\tau]$. To calculate $G$ we need the first and second derivatives of $C^-$ with respect to $\lambda_1$ and $\lambda_2$. These are
\begin{eqnarray}
\frac{\partial C^-}{\partial\lambda_j}  & = & \frac{C^-}{2\lambda_j}, \\
\frac{\partial^2 C^-}{\partial\lambda_j^2} & = & - \frac{C^-}{4\lambda_j^2}, \\
\frac{\partial^2 C^-}{\partial\lambda_1\partial\lambda_2} & = &  \frac{C^-}{4\lambda_1\lambda_2}.
\end{eqnarray}
The expression we get for $G$ this time is
\begin{eqnarray}
   G(X) & = &  (X_{11} - X_{22})^2 C^-  \nonumber \\
     & & + 4 \left( |X_{10}|^2 + |X_{20}|^2 + |X_{21}|^2 \right) C^- .
\end{eqnarray}
We therefore need to find the $U$ that maximizes
\begin{eqnarray*}
   F(X) & = &  (X_{11} - X_{22})^2 + 4 \left( |X_{10}|^2 + |X_{20}|^2 + |X_{21}|^2 \right) .
\end{eqnarray*}
Once again we perform this optimization numerically using Matlab's fminsearch. It turns out that there are many unitaries that maximize $F(X)$, and the unitaries that give the $X$'s defined by Eqs.~(\ref{Xmax}) and (\ref{Xmax2}) both do so. So we have
\begin{equation}
  \max_{U} G(X) =  4 a^2 C^- . 
\end{equation}
This is exactly the time derivative of $C^-$, so the cost function does satisfy the HJB equation on the interval $[T-\tau,T]$.

Our final task is to determine whether the cost function is a
viscosity solution to the HJB equation at $t = T-\tau$. To do this
we would usually proceed to calculate the super and subderivatives,
which would involve expanding $C^+$ and $C^-$ in their Taylor series
as describe in Sec.~\ref{VS}. However, it turns out that in this
case we can take a shortcut. First we note that the protocol
involves switching between two measurement operators, so the cost
function must be a solution to the HJB equations for both. The two
HJB equations are
\begin{eqnarray}
  \frac{\partial C}{\partial t} & = & 8 |X_{10}|^2 \lambda_1 \frac{\partial C}{\partial \lambda_1} ,  \label{HJBf1} \\
  \frac{\partial C}{\partial t} & = & 8 |X_{20}|^2 \lambda_2 \frac{\partial C}{\partial \lambda_2} .   \label{HJBf2}
\end{eqnarray}
Because these do not contain the second derivatives, we do not need to calculate the $Q$ part of the super and subderivatives to determine if $C$ is a viscosity solution. We only need the first order parts, $q$ and $\bp$. But since the first derivatives of $C$ are continuous at $t = T-\tau$, $q$ and $\bp$ are exactly these derivatives~\footnote{Strictly speaking this is only true if the super and subderivatives are nonempty, but if they are empty then $C$ is automatically a viscosity solution to the HJB equation at $t = T-\tau$.}. So all we need to do is to substitute $C^+$ (or $C^-$) into Eqs.~(\ref{HJBf1}) and (\ref{HJBf2}), and check that it is a solution to both. Indeed it is, and thus $C$ is a viscosity solution of the HJB equation, and we can conclude that the suggested protocol is optimal.

\section{Finding Optimal Protocols}
\label{findop}
The two verification theorems we have described above also provide a method to search for optimal protocols. To see how this works we return to the Hamilton-Jacobi-Bellman equation, which is
\begin{equation}
  \frac{\partial C}{\partial t} = \max_{\mathbf{v}} \left[ G\left( \mathbf{x}, \mathbf{v}, \frac{\partial C}{\partial \mathbf{x}}, \frac{\partial^2 C}{\partial \mathbf{x}^2} \right) \right] .
  \label{HJB_again}
\end{equation}
Recall that $C(\mathbf{x},t)$ is the total average cost
over the remaining time interval, $[t,T]$, given that the system has
state $\mathbf{x}$ at time $t$. The elements of the  vector
$\mathbf{v}$, in general being functions of the state $\mathbf{x}$
and time $t$, are the parameters that give the control protocol. The
evolution of the system, $\mathbf{x}(t)$, and the cost function
$C(\mathbf{x},t)$, are determined by the choice of
$\mathbf{v}$. If the pair $\mathbf{v}$ and $C$ solve the HJB
equation when substituted into it, then $\mathbf{v}$ is an optimal
protocol.

The fact that allows us to use the HJB equation to search for optimal protocols, is that {\em every} pair of $\mathbf{v}$ and $C$ satisfies the equation
\begin{equation}
  \frac{\partial C}{\partial t} =  G\left( \mathbf{x}, \mathbf{v}, \frac{\partial C}{\partial \mathbf{x}}, \frac{\partial^2 C}{\partial \mathbf{x}^2} \right) .
  \label{relaxHJB}
\end{equation}
However, $G$ will only be maximized if $\mathbf{v}$ is an optimal
protocol. This means that we can find an optimal protocol by
searching over the allowed space of controls $\mathbf{v}$ for the
$\mathbf{v}$ that maximizes $G$. Note that if $\mathbf{v}$ is
time-dependent, then we will have to search over all functions
$\mathbf{v}$ for one that maximizes $G$ at {\em every} time $t$.
This can certainly be a challenging task -- nevertheless, it does
provide a systematic procedure. Further, the problem simplifies in
the following way: one can perform the search by starting close to
the final time, and stepping backwards. That is, solving for the
optimal $\mathbf{v}$ in the small time interval $[T-\Delta t, T]$,
for all states $x$ at the start of this interval. Then, with
$\mathbf{v}(x,t)$ for that interval fixed at the optimal result
obtained, solving for the interval $[T-2\Delta t, T-\Delta]$, and
so-on moving backwards. Because of the form of the cost,
Eq.(\ref{cost}), this backwards-in-time procedure will find a
globally optimal $\mathbf{v}$ for the whole interval. (This point is
discussed in most control texts that include the Bellman equation - see
e.g.~\cite{Whittle,KSeq}.)

If $\mathbf{v}$ is not a function of time, but only of the state, $\mathbf{x}$, then the task is significantly easier, since not only is the search space reduced, but it usually means that if $G$ is maximized at a single time, it is maximized at all times, so we need only evaluate $G$ at a single time. If one can obtain an analytic solution to the equations of motion for every $\mathbf{v}$, then one can obtain an analytic expression for $G$ in terms of $\mathbf{v}$. In this case numerical minimization is likely to be easy.

\section{An Alternative Form for the Cost: Time-Optimal Control}
\label{OCO}

So far we have taken the cost to be a function of the dynamical
variables, integrated over time, including a separate contribution
at the final time. There is another useful form for the cost
function that allows us to use the exactly the same verification
procedures. This alternative cost is what we use if we want to
minimize the {\em time} taken to reach a particular event. This kind
of control problem is called {\em time-optimal control}. In this
case we define a function $h(\mathbf{x}(t),t)$ of the dynamical
variables (and perhaps of time), and the goal is to minimize the
time taken for $h$ to cross a fixed value $h_{\mbox{\scriptsize
c}}$, called the {\em threshold}. The cost is then defined as the
expectation value of the time remaining before $h$ crosses the
threshold. Everything else is the same as before, the only change is
this form of the cost. As before, the cost function,
$C(\mathbf{x},t)$, is the average value of the cost from time
$t$ until we finish, given that the state at time $t$ is
$\mathbf{x}$. That means that $C(\mathbf{x},t)$ is the average
value of time it will take to cross the threshold, given that the
current time is $t$ and current state is $\mathbf{x}$. Thus if we
define $\mathbf{x}_{\mbox{\scriptsize c}}$ as the state (or set of
states) for which $h = h_{\mbox{\scriptsize c}}$, then
$C(\mathbf{x}_{\mbox{\scriptsize c}},t) = 0$.

The only change in the verification theorem is that the HJB equation changes a little. It is now~\cite{Fleming75}
\begin{equation}
  \frac{\partial C}{\partial t} = \max_{\mathbf{v}} \left[ G\left( \mathbf{x}, \mathbf{v}, \frac{\partial C}{\partial \mathbf{x}}, \frac{\partial^2 C}{\partial \mathbf{x}^2} \right) \right] - 1,
  \label{HJBtoc}
\end{equation}
where $G$ is the same as before, except that it no longer contains the function $L$:
\begin{equation}
  G = - \frac{1}{2} \mbox{Tr} \left[ B^{\dagger}(t,\mathbf{x},\mathbf{v}) \frac{\partial^2 C}{\partial \mathbf{x}^2} B(t,\mathbf{x},\mathbf{v})  \right]  -  \mathbf{A} \bcdot \frac{\partial C}{\partial \mathbf{x}} .
\end{equation}
Note that the HJB equation does not contain either the function $h(\mathbf{x}(t),t)$, or the threshold value $h_{\mbox{\scriptsize c}}$~\footnote{The terminal condition on the time-optimal HJB equation is $C(\mathbf{x}_{\mbox{\scriptsize c}},t) = 0$. This is, of course, satisfied by the cost function, $C$, since it is part of its definition.}. These have already played their role by determining $C$.

\section{Conclusion}
\label{conc}

We have shown how to use verification theorems to determine whether a given feedback protocol is optimal. The procedure is quite straightforward, and involves: 1.\ solving the equations of motion to obtain the cost function,  2.\ checking two simple continuity conditions on the cost function, 3.\ checking that the cost function is a solution of the HJB equation, and 4.\ checking that the protocol maximizes the RHS of the HJB equation.

If the cost function does not satisfy the continuity conditions (step 2 above), then there is an enhanced verification procedure that can be used. This is essentially the same as the previous procedure, but one checks instead that the cost function is a {\em viscosity} solution of the HJB equation.

We have also described how the HJB equation can be used, at least in principle, to find optimal protocols. For non-trivial problems this will usually, but not always, require a numerical implementation, and if so is likely to be numerically intensive. What we have not yet done is to show how the HJB equation is derived. This is actually quite straightforward, and we give this derivation in the Appendix.

\section{Appendix} 

\subsection{The Hamilton-Jacobi-Bellman Equation}
\label{app}

To begin with we define the {\em value function}, $V(\bx,t)$, as the
minimum of the cost function over all possible control protocols:
\begin{equation}
  V(\mathbf{x}(t),t) = \min_{\bu} C_{\mathbf{u}}(\mathbf{x}(t),t) , 
\end{equation}
where we have added the subscript $\mathbf{u}$ to the cost function to make explicit the fact that it depends on the choice of the control, $\mathbf{u}(\mathbf{x}(t),t)$. Recall that the cost function, $C_{\mathbf{u}}(\mathbf{x}(t),t)$, is the average value of the cost that will be incurred from time $t$ until the final time. (The cost function is defined in Section~\ref{CVT}). The starting point for the HJB equation is the fact that the value function satisfies the recursion relation
\begin{equation}
  V(\mathbf{x}(t),t) = \min_{\bu(\bx,t)} \left\langle \int_t^{t^{'}} \!\!\! L \; ds +V(\mathbf{x}(t'),t') \right\rangle_{\mbox{\scriptsize n}} ,
  \label{recur2}
\end{equation}
where $L = L(\mathbf{u}[\mathbf{x}(s),s],\mathbf{x}(s),s)$ is the instantaneous cost incurred at time $s$. This recursion relation follows immediately from two facts. The first is that any protocol that gives the optimal control from time $0$ to $T$, must also give the optimal control from any time $t' > 0$ to $T$. Recall that the optimal protocol tells us what control to apply at every time $t$, for every state $\mathbf{x}(t)$ that the system could be in at that time. Thus if the protocol $u$ minimizes $C_{\mathbf{u}}(\mathbf{x}(t),t)$, then it also minimizes $C_{\mathbf{u}}(\mathbf{x}(t'),t')$ for $t'>t$. The second fact is that the total cost is merely a sum (integral) over the separate costs $L(t)$ at each time $t$. Because of this the total average cost incurred from time $t$ onwards is the sum of the cost incurred from time $t$ to $t'$, and the cost incurred from $t'$ onwards. That is 
\begin{equation}
  C_{\mathbf{u}}(\mathbf{x}(t),t) = \left\langle \int_t^{t^{'}} \!\!\! L \; ds +C_{\mathbf{u}}(\mathbf{x}(t'),t') \right\rangle_{\mbox{\scriptsize n}} .
  \label{recur2}
\end{equation}
Adding to this the fact that the optimal protocol $\mathbf{u}$ minimizes both $C_{\mathbf{u}}(\mathbf{x}(t),t)$ and $C_{\mathbf{u}}(\mathbf{x}(t'),t')$, we obtain the recursion relation for $V$ (Eq.(\ref{recur2})). 


The HJB equation can be extracted from Eq.(\ref{recur2}) by
setting $t' = t+dt$ and applying Ito's rule, $dW^2 = dt$. Note that this means
we must keep all the differentials up to the second order. To begin with we
have
\begin{equation}
   V(\mathbf{x},t) = \min_{\bu(\bx,t)} \left\langle \int_t^{t+dt} \!\!\!\!\!\!\! L \; ds \right\rangle_{\mbox{\scriptsize n}}
    \left\langle V(\mathbf{x}(t+dt),t+dt) \right\rangle_{\mbox{\scriptsize n}} .
    \label{Vde}
\end{equation}
We now substitute into this the Taylor expansion for $V(\mathbf{x}+d\mathbf{x},t+dt)$, being
\begin{eqnarray*}
   V(\mathbf{x}+d\mathbf{x},t+dt) & = & V(t,\mathbf{x}) + dt \frac{\partial V}{\partial t}  +d\mathbf{x} \bcdot \frac{\partial V}{\partial \mathbf{x}} \\
             &  & +d\mathbf{x}^{\dagger} \bcdot
\frac{\partial^2 V}{\partial \mathbf{x}^2} \bcdot d\mathbf{x} ,
\end{eqnarray*}
and the equation of motion for the system, (Eq.(\ref{sys-dy})), being
\begin{equation}
\mathbf{x}+d\mathbf{x}=\mathbf{x}+\mathbf{A}(t,\mathbf{x},\mathbf{u}(\mathbf{x},t))
dt + B(t,\mathbf{x},\mathbf{u}(\mathbf{x},t)) \mathbf{dW} .
\end{equation}
The result is the HJB equation
 \begin{eqnarray}
   \frac{\partial V}{\partial t}& = & \max_{\bu(\bx,t)}  \left\{ -\frac{1}{2} \mbox{Tr} \left[ B^{\dagger}(t,\mathbf{x},\mathbf{u}) \frac{\partial^2 C}{\partial \mathbf{x}^2} B(t,\mathbf{x},\mathbf{u})  \right]  \right. \nonumber \\
             &  &  \left. -\mathbf{A}  \bcdot \frac{\partial C}{\partial
             \mathbf{x}}-L(t,\mathbf{x},\mathbf{u})  \right\} .
\end{eqnarray}
Notice that in the last step we used the fact that
$-\mbox{min}(f)=\mbox{max}(-f)$ for any function $f$.
A rigorous derivation of the HJB equation can be
found in many stochastic control textbooks, an example being Ref.~\cite{Fleming75}.


\begin{thebibliography}{90}
\providecommand{\natexlab}[1]{#1}

\bibitem[1]{JacobsSteck06}
K. Jacobs and D.A. Steck, {\itshape A Straightforward Introduction to
  Continuous Quantum Measurement}, Contemp. Phys. 47 (2006), pp. 279--303.

\bibitem[2]{Brun02}
T.A. Brun, {\itshape A simple model of quantum trajectories}, {Am.\ J.\ Phys.}
  70 (2002), pp. 719--737.

\bibitem[3]{Belavkin87}
V.P. Belavkin, {\itshape Non-demolition measurement and control in quantum
  dynamical systems}, A.~Blaquiere, S.~Diner and G.~Lochak,  eds., in {\em
  Information, Complexity and Control in Quantum Physics. Proceedings of the
  4th International Seminar on Mathematical Theory of Dynamical Systems and
  Microphysics}, Udine, 4--13 Sept. 1985,   Springer-Verlag, New York, 1987,
  pp. 331--336.

\bibitem[4]{Belavkin83}
---{}---{}---, {\itshape Theory of the control of observable quantum systems},
  Automatica and Remote Control 44 (1983), pp. 178--188.

\bibitem[5]{Wiseman95}
H.M. Wiseman, {\itshape Adaptive phase measurements of optical modes: Going
  beyond the marginal Q distribution}, Phys.\ Rev.\ Lett. 75 (1995), pp.
  4587--4590.

\bibitem[6]{Wiseman95b}
---{}---{}---, {\itshape Using feedback to eliminate back-action in quantum
  measurements}, Phys. Rev. A 51 (1995), pp. 2459--2468.

\bibitem[7]{Yanagisawa98}
M. Yanagisawa and H. Kimura, {\itshape A Control Problem for Gaussian States},
  in  {\itshape Learning, Control and Hybrid Systems, Lecture Notes in Control
  and Information Sciences}, Y.~Yamamoto and S.~Hara,  eds.,
  Springer-Verlag, New York, 1998, pp. 249--313.

\bibitem[8]{DJ}
A.C. Doherty and K. Jacobs, {\itshape Feedback control of quantum systems using
  continuous state estimation}, Phys.\ Rev.\ A 60 (1999), pp. 2700--2711.

\bibitem[9]{Belavkin}
V.P. Belavkin, {\itshape Measurement, filtering and control in quantum open
  dynamical systems}, Rep. Math. Phys. 43 (1999), pp. 405--425.

\bibitem[10]{DHJMT}
A.C. Doherty, S. Habib, K. Jacobs, H. Mabuchi, and S.M. Tan, {\itshape Quantum
  feedback control and classical control theory}, Phys. Rev. A 62 (2000), p.
  012105.

\bibitem[11]{DJJ}
A.C. Doherty, K. Jacobs, and G. Jungman, {\itshape Information, disturbance,
  and Hamiltonian quantum feedback control}, Phys. Rev. A 63 (2001), p. 062306.

\bibitem[12]{Thomsen02}
L.K. Thomsen, S. Mancini, and H.M. Wiseman, {\itshape Spin squeezing via
  quantum feedback}, Phys. Rev. A 65 (2002), p. 061801.

\bibitem[13]{Yanagisawa03a}
M. Yanagisawa and H. Kimura, {\itshape Transfer function approach to quantum
  control-part I: Dynamics of quantum feedback systems}, IEEE Trans. Automat.
  Contr. 48 (2003), pp. 2107--2120.

\bibitem[14]{Yanagisawa03b}
---{}---{}---, {\itshape Transfer function approach to quantum Control-Part II:
  Control concepts and applications}, IEEE Trans. Automat. Contr. 48 (2003),
  pp. 2121--2132.

\bibitem[15]{Geremia03}
J. Geremia, J.K. Stockton, A.C. Doherty, and H. Mabuchi, {\itshape Quantum
  Kalman Filtering and the Heisenberg Limit in Atomic Magnetometry}, Phys. Rev.
  Lett. 91 (2003), p. 250801.

\bibitem[16]{Bouten04}
L. Bouten, M. Guta, and H. Maassen, {\itshape Stochastic Schrodinger
  equations}, J. Phys. A 37 (2004), pp. 3189--3209.

\bibitem[17]{James04}
M.R. James, {\itshape Risk-Sensitive Optimal Control of Quantum Systems}, Phys.
  Rev. A 69 (2004), p. 032108.

\bibitem[18]{Bouten05}
L. Bouten, S. Edwards, and V.P. Belavkin, {\itshape Bellman equations for
  optimal feedback control of qubit states}, J. Phys. B 38 (2005), pp.
  151--160.

\bibitem[19]{vanHandel05}
R. vanHandel, J.K. Stockton, and H. Mabuchi, {\itshape Feedback control of
  quantum state reduction}, IEEE T. Automat. Contr. 50 (2005), pp. 768--780.

\bibitem[20]{Gough05}
J. Gough, V.P. Belavkin, and O.G. Smolyanov, {\itshape Hamilton-Jacobi-Bellman
  equations for Quantum Filtering and Control}, J. Opt. B: Quantum Semiclass.
  Opt. 7 (2005), pp. S237--S244.

\bibitem[21]{Bouten05e}
L. Bouten and R. vanHandel, {\itshape On the separation principle of quantum
  control}, Eprint: arXiv:math-ph/0511021 .

\bibitem[22]{Wiseman05}
H.M. Wiseman and A.C. Doherty, {\itshape Optimal Unravellings for Feedback
  Control in Linear Quantum Systems}, Phys.\ Rev.\ Lett. 94 (2005), p. 070405.

\bibitem[23]{Mirrahimi05-1}
M. Mirrahimi, P. Rouchon, and G. Turinici, {\itshape Lyapunov control of
  bilinear Schrodinger equations}, Automatica 41 (2005), pp. 1987--1994.

\bibitem[24]{Mirrahimi05-2}
M. Mirrahimi, G. Turinici, and P. Rouchon, {\itshape Reference trajectory
  tracking for locally designed coherent quantum controls}, J. Phys. Chem. A
  109 (2005), pp. 2631--2637.

\bibitem[25]{DHelon06}
C. D'Helon and M.R. James, {\itshape Stability, gain, and robustness in quantum
  feedback networks}, Phys.\ Rev.\ A 73 (2006), p. 053803.

\bibitem[26]{Dong06}
D.Y. Dong, C.L. Chen, Z.H. Chen, and C.B. Zhang, {\itshape Estimation-based
  information acquisition in quantum feedback control}, Dynam. Cont. Dis. Ser.
  B 13 (2006), pp. 1204--1208.

\bibitem[27]{Combes06}
J. Combes and K. Jacobs, {\itshape Rapid state-reduction of quantum systems
  using feedback control}, Phys. Rev. Lett. 96 (2006), p. 010504.

\bibitem[28]{Altafini06}
C. Altafini, {\itshape Homogeneous polynomial forms for simultaneous
  stabilizability of families of linear control systems: A tensor product
  approach}, IEEE T. Automat. Contr. 51 (2006), pp. 1566--1571.

\bibitem[29]{Altafini07}
---{}---{}---, {\itshape Decoherence Control by Tracking a Hamiltonian
  Reference Molecule}, Quant. Inf. Proc. 6 (2007), pp. 9--36.

\bibitem[30]{Altafini07b}
---{}---{}---, {\itshape Feedback stabilization of isospectral control systems
  on complex flag manifolds: Application to quantum ensembles}, IEEE T.
  Automat. Contr. 52 (2007), pp. 2019--2028.

\bibitem[31]{Ticozzi06}
F. Ticozzi and L. Viola, {\itshape Single-bit feedback and quantum-dynamical
  decoupling}, Phys. Rev. A 74 (2006), p. 052328.

\bibitem[32]{Pavon06}
M. Pavon and F. Ticozzi, {\itshape On entropy production for controlled
  Markovian evolution}, J. Math. Phys. 47 (2006), p. 063301.

\bibitem[33]{vanHandel07}
M. Mirrahimi and R. vanHandel, {\itshape Stabilizing feedback controls for
  quantum systems}, SIAM J. Control Optim. 46 (2007), pp. 445--467.

\bibitem[34]{Wang07}
S.K. Wang, J.S. Jin, and X.Q. Li, {\itshape Continuous weak measurement and
  feedback control of a solid-state charge qubit: A physical unravelling of
  non-Lindblad master equation}, Phys. Rev. B 75 (2007), p. 155304.

\bibitem[35]{Jacobs07c}
K. Jacobs and A.P. Lund, {\itshape Feedback Control of Nonlinear Quantum
  Systems: A Rule of Thumb}, Phys. Rev. Lett. 99 (2007), p. 020501.

\bibitem[36]{Ticozzi08}
F. Ticozzi and L. Viola, {\itshape Quantum Markovian Subsystems: Invariance,
  Attractivity, and Control}, IEEE T. Automat. Contr. 53 (2008), pp.
  2048--2063.

\bibitem[37]{Gough08}
J. Gough and M.R. James, {\itshape The Series Product and Its Application to
  Quantum Feedforward and Feedback Networks}, Eprint: arXiv:0707.0048  (2007).

\bibitem[38]{James08}
M.R. James and J. Gough, {\itshape Quantum Dissipative Systems and Feedback
  Control Design by Interconnection}, Eprint: arXiv:0707.1074  (2007).

\bibitem[39]{Nurdin08}
H.I. Nurdin, M.R. James, and I.R. Petersen, {\itshape Coherent quantum LQG
  control}, Eprint: arXiv:0711.2551  (2007).

\bibitem[40]{Belavkin08}
V.P. Belavkin, A. Negretti, and K. Molmer, {\itshape Dynamical programming of
  continuously observed quantum systems}, Eprint: arXiv:0805.4741  (2008).

\bibitem[41]{Smith02}
W.P. Smith, J.E. Reiner, L.A. Orozco, S. Kuhr, and H.M. Wiseman, {\itshape
  Capture and Release of a Conditional State of a Cavity QED System by Quantum
  Feedback}, Phys.\ Rev.\ Lett. 89 (2002), p. 133601.

\bibitem[42]{Armen02}
M.A. Armen, J.K. Au, J.K. Stockton, A.C. Doherty, and H. Mabuchi, {\itshape
  Adaptive Homodyne Measurement of Optical Phase}, Phys.\ Rev.\ Lett. 89
  (2002), p. 133602.

\bibitem[43]{Cook07}
R.L. Cook, P.J. Martin, and J.M. Geremia, {\itshape Optical coherent state
  discrimination using a closed-loop quantum measurement}, Nature 446 (1992),
  pp. 774--777.

\bibitem[44]{Higgins07}
B.L. Higgins, D.W. Berry, S.D. Bartlett, H.M. Wiseman, and G.J. Pryde,
  {\itshape Entanglement-free Heisenberg-limited phase estimation}, Nature 450
  (2007), pp. 393--396.

\bibitem[45]{Naik06}
A. Naik, O. Buu, M.D. LaHaye, A.D. Armour, A.A. Clerk, M.P. Blencowe, and K.C.
  Schwab, {\itshape Cooling a nanomechanical resonator with quantum
  back-action}, Nature 443 (2006), pp. 193--196.

\bibitem[46]{Niskanen07}
A.O. Niskanen, K. Harrabi, F. Yoshihara, Y. Nakamura, S. Lloyd, and J.S. Tsai,
  {\itshape Quantum Coherent Tunable Coupling of Superconducting Qubits},
  Science 316 (2007), pp. 723--726.

\bibitem[47]{Houck07}
A.A. Houck, D.I. Schuster, J.M. Gambetta, J.A. Schreier, B.R. Johnson, J.M. Chow, L. Frunzio, J. Majer, M.H. Devoret, S.M. Girvin, and R.J. Schoelkopf, {\itshape Generating single microwave photons in a circuit}, Nature 449 (2007), pp. 328--331.

\bibitem[48]{Schuster07}
D.I. Schuster, A.A. Houck, J.A. Schreier, A. Wallraff, J.M. Gambetta, A. Blais, L. Frunzio, J. Majer, B. Johnson, M.H. Devoret, S.M. Girvin and R.J. Schoelkopf, {\itshape Resolving photon number states in a superconducting circuit}, Nature 445 (2007), pp. 515--518.

\bibitem[49]{Plantenberg07}
J.H. Plantenberg, P.C. deGroot, C.J.P.M. Harmans, and J.E. Mooij, {\itshape
  Demonstration of controlled-NOT quantum gates on a pair of superconducting
  quantum bits}, Nature 447 (2007), pp. 836--839.

\bibitem[50]{Ahn02}
C. Ahn, A.C. Doherty, and A.J. Landahl, {\itshape Continuous quantum error
  correction via quantum feedback control}, Phys. Rev. A 65 (2002), p. 042301.

\bibitem[51]{rapidP}
K. Jacobs, {\itshape How to project qubits faster using quantum feedback},
  Phys. Rev. A 67 (2003), p. 030301(R).

\bibitem[52]{Hopkins03}
A. Hopkins, K. Jacobs, S. Habib, and K. Schwab, {\itshape Feedback cooling of a
  nanomechanical resonator}, Phys. Rev. B 68 (2003), p. 235328.

\bibitem[53]{Geremia04}
J.M. Geremia, {\itshape Distinguishing between optical coherent states with
  imperfect detection}, Phys. Rev. A 70 (2004), p. 062303.

\bibitem[54]{Ralph04}
J.F. Ralph, E.J. Griffith, T.D. Clark, and M.J. Everitt, {\itshape Guidance and
  Control in a Josephson Charge Qubit}, Phys. Rev. B 70 (2004), p. 214521.

\bibitem[55]{Sarovar04}
M. Sarovar, C. Ahn, K. Jacobs, and G.J. Milburn, {\itshape Practical scheme for
  error control using feedback}, Phys.\ Rev.\ A 69 (2004), p. 052324.

\bibitem[56]{Steck04}
D. Steck, K. Jacobs, H. Mabuchi, T. Bhattacharya, and S. Habib, {\itshape
  Quantum Feedback Control of Atomic Motion in an Optical Cavity}, Phys. Rev.
  Lett. 92 (2004), p. 223004.

\bibitem[57]{Steixner05}
V. Steixner, P. Rabl, and P. Zoller, {\itshape Quantum feedback cooling of a
  single trapped ion in front of a mirror}, Phys. Rev. A 72 (2005), p. 043826.

\bibitem[58]{Yamamoto05}
N. Yamamoto, {\itshape Parametrization of the feedback Hamiltonian realizing a
  pure steady state}, Phys. Rev. A 72 (2005), p. 024104.

\bibitem[59]{Yanagisawa06}
M. Yanagisawa, {\itshape Quantum Feedback Control for Deterministic Entangled
  Photon Generation}, Phys. Rev. Lett. 97 (2006), p. 190201.

\bibitem[60]{Wiseman06x}
H.M. Wiseman and J.F. Ralph, {\itshape Reconsidering rapid qubit purification
  by feedback}, New. J. Phys 8 (2006), p.~90.

\bibitem[61]{Bushev06}
P. Bushev, D. Rotter, A. Wilson, F. Dubin, C. Becher, J. Eschner, R. Blatt, V.
  Steixner, P. Rabl, and P. Zoller, {\itshape Feedback Cooling of a Single
  Trapped Ion}, {Phys.\ Rev.\ Lett.} 96 (2006), p. 043003.

\bibitem[62]{Matsukevich06}
D.N. Matsukevich, T. Chaneliere, S.D. Jenkins, S.Y. Lan, T.A.B. Kennedy, and A.
  Kuzmich, {\itshape Deterministic single photons via conditional quantum
  evolution}, Phys. Rev. Lett. 97 (2006), p. 013601.

\bibitem[63]{Mancini06}
L. Tian, {\itshape Markovian feedback to control continuous variable
  entanglement}, Phys. Rev. A 73 (2006), p. 010304(R).

\bibitem[64]{Katz07}
G. Katz, M.A. Ratner, and R. Kosloff, {\itshape Decoherence Control by Tracking
  a Hamiltonian Reference Molecule}, Phys. Rev. Lett. 98 (2007), p. 203006.

\bibitem[65]{Tian07}
L. Tian, {\itshape Correcting Low-Frequency Noise with Continuous Measurement},
  Phys. Rev. Lett. 98 (2007), p. 153602.

\bibitem[66]{Jacobs07}
K. Jacobs, {\itshape Feedback control for communication with non-orthogonal
  states}, Quant. Information Comp. 7 (2007), pp. 127--138.

\bibitem[67]{Hill07}
C. Hill and J.F. Ralph, {\itshape Weak measurement and rapid state reduction in
  entangled bipartite quantum systems}, New J. Phys 9 (2007), p. 151.

\bibitem[68]{Carvalho07}
A.R.R. Carvalho and J.J. Hope, {\itshape Stabilizing entanglement by
  quantum-jump-based feedback}, Phys. Rev. A 76 (2007), p. 010301.

\bibitem[69]{Wilson07}
S.D. Wilson, A.R.P. Carvalho, J.J. Hope, and M.R. James, {\itshape Effects of
  measurement backaction in the stabilization of a Bose-Einstein condensate
  through feedback}, Phys. Rev. A 76 (2007), p. 013610.

\bibitem[70]{Yamamoto07}
N. Yamamoto, K. Tsumura, and S. Hara, {\itshape Feedback control of quantum
  entanglement in a two-spin system}, Automatica 43 (2007), pp. 981--992.

\bibitem[71]{Bradley08}
B.A. Chase, A.J. Landahl, and J.M. Geremia, {\itshape Efficient feedback
  controllers for continuous-time quantum error correction}, Phys. Rev. A 77
  (2008), p. 032304.

\bibitem[72]{Combes08}
J. Combes, H.M. Wiseman, and K. Jacobs, {\itshape Rapid Measurement of Quantum
  Systems Using Feedback Control}, Phys. Rev. Lett. 100 (2008), p. 160503.

\bibitem[73]{Wiseman08}
H.M. Wiseman and L. Bouten, {\itshape Optimality of feedback control strategies
  for qubit purification}, Quant. Inform. Proc. 7 (2008), pp. 71--83.

\bibitem[74]{Sagawa08}
T. Sagawa and M. Ueda, {\itshape Second Law of Thermodynamics with Discrete
  Quantum Feedback Control}, Phys. Rev. Lett. 100 (2008), p. 080403.

\bibitem[75]{FJ}
C.A. Fuchs and K. Jacobs, {\itshape Information-tradeoff relations for
  finite-strength quantum measurements}, Phys. Rev. A 63 (2001), p. 062305.

\bibitem[76]{Fleming75}
W.H. Fleming and R.W. Rishel {\itshape Deterministic and Stochastic Optimal
  Control},    Springer-Verlag, New York, 1975.

\bibitem[77]{Boltianskii66}
V.G. Boltyanskii, {\itshape Sufficient conditions for optimality and the
  justification of the dynamic programming method}, SIAM J. Control 4 (1966),
  pp. 326--361.

\bibitem[78]{Vinter80}
R.B. Vinter and R.M. Lewis, {\itshape Verification theorem which provides
  provides a necessary and sufficient condition for optimality}, IEEE T.
  Automat. Contr. 25 (1980), pp. 84--89.

\bibitem[79]{Zhou97}
X.Y. Zhou, J. Yong, and X. Li, {\itshape Stochastic verification theorems
  within the framework of Viscosity Solutions}, SIAM J. Cont. Optim. 35 (1997),
  pp. 243--253.

\bibitem[80]{Gozzi05}
F. Gozzi, A. Swiech, and X.Y. Zhou, {\itshape A corrected proof of the
  stochastic verification theorem within the framework of viscosity solutions},
  SIAM J. Cont. Optim. 43 (2005), pp. 2009--2019.

\bibitem[81]{Crandall92}
M.G. Crandall, H. Ishii, and P.L. Lions, {\itshape User's guide to viscosity
  solutions of second order partial differential equations}, B. Am. Math. Soc.
  27 (1992), pp. 1--67.

\bibitem[82]{Shabani08}
A. Shabani and K. Jacobs, {\itshape Locally Optimal Control of Quantum Systems
  with Strong Feedback}, Phys. Rev. Lett. 101 (2008), 230403.

\bibitem[83]{WienerIntroPaper}
D.T. Gillespie, {\itshape The mathematics of Brownian motion and Johnson
  noise}, Am. J. Phys. 64 (1996), pp. 225--240.

\bibitem[84]{KJPhD}
K. Jacobs, {\itshape Topics in Quantum Measurement and Quantum Noise}, Ph.D.
  diss., Imperial College, London, 1998.

\bibitem[85]{FuchsPhD}
C.A. Fuchs, {\itshape Distinguishability and Accessible Information in Quantum
  Theory}, Ph.D. diss., UNM, Albuquerque, 1995.

\bibitem[86]{Fuchs98}
---{}---{}---, {\itshape Information gain vs. state disturbance in quantum
  theory}, Fort. der Phys. 46 (1998), pp. 535--565.

\bibitem[87]{Vedral97}
V. Vedral, M.B. Plenio, K. Jacobs, and P.L. Knight, {\itshape Statistical
  inference, distinguishability of quantum states, and quantum entanglement},
  Phys. Rev. A 56 (1997), pp. 4452--4455.

\bibitem[88]{Dita03}
P. Di\c{t}\v{a}, {\itshape Factorization of unitary matrices}, J. Phys. A 36
  (2003), pp. 2781--2789.

\bibitem[89]{Whittle}
P. Whittle {\itshape Optimal Control},    Wiley, Chichester, 1996.

\bibitem[90]{KSeq}
P.S. Maybeck {\itshape Stochastic Models, Estimation and Control},   Vol. I and
  II,   Academic Press, New York, 1982.

\end{thebibliography}

\end{document}